\documentclass[10pt,conference]{IEEEtran}
\usepackage{amsmath,amsfonts,amsthm}
\usepackage{algorithm}
\usepackage{algpseudocode}
\usepackage{array}
\usepackage{algorithmicx}
\usepackage[caption=false,font=normalsize,labelfont=sf,textfont=sf]{subfig}
\usepackage{textcomp}
\usepackage{verbatim}
\usepackage{stfloats}
\usepackage{url}
\usepackage{verbatim}
\usepackage{graphicx}
\usepackage{cite}
\usepackage{eurosym}
\usepackage{xurl}
\theoremstyle{plain}

\usepackage{longtable}
\usepackage{tikz}
\usepackage{quantikz}
\usetikzlibrary{positioning, arrows, fit, automata}
\definecolor{Lightblue}{RGB}{100,100,230}
\definecolor{Mediumblue}{RGB}{20,20,170}
\definecolor{Darkblue}{RGB}{20,20,100}

\usepackage{pgfplots}
\pgfplotsset{compat=1.18}
\usepgfplotslibrary{groupplots}
\usepgfplotslibrary{statistics}

\usepackage{bbm} 

\begin{document}


\definecolor{plasmaFiveOne}{RGB}{238, 123, 81}
\definecolor{plasmaFiveTwo}{RGB}{205, 73, 119}
\definecolor{plasmaFiveThree}{RGB}{157, 24, 157}
\definecolor{plasmaFiveFour}{RGB}{94, 1, 166}
\definecolor{plasmaFiveFive}{RGB}{13, 8, 135}

\newcommand{\colormapOne}{plasmaFiverOne}
\newcommand{\colormapTwo}{plasmaFiveTwo}
\newcommand{\colormapThree}{plasmaFiveThree}
\newcommand{\colormapFour}{plasmaFiveFour}
\newcommand{\colormapFive}{plasmaFiveFive}

\newcommand{\colormapOneFill}{plasmaFiverOne!20}
\newcommand{\colormapTwoFill}{plasmaFiveTwo!20}
\newcommand{\colormapThreeFill}{plasmaFiveThree!20}
\newcommand{\colormapFourFill}{plasmaFiveFour!20}
\newcommand{\colormapFiveFill}{plasmaFiveFive!20}

\newcommand{\marker}{diamond}
\newcommand{\markerfig}{diamond*}

\title{From Cables to Qubits: A Decomposed Variational Quantum Optimization Pipeline}

\author{
  \IEEEauthorblockN{
    Paul-Niklas Ken Kandora\IEEEauthorrefmark{1}\thanks{Corresponding author: paul-niklas.kandora@kit.edu},
    Adrian Asmund Fessler\IEEEauthorrefmark{1},
    Robert Fabian Lindermann\IEEEauthorrefmark{1},
    Phil Arnold\IEEEauthorrefmark{2},\\[-0.2ex]%
    Andreas Hempel\IEEEauthorrefmark{2},
    Steffen Rebennack\IEEEauthorrefmark{1}
  }
  \\
  \IEEEauthorblockA{
    \IEEEauthorrefmark{1}Karlsruhe Institute of Technology, Karlsruhe, Germany\\
    \sloppy
    \texttt{\{paul-niklas.kandora,\allowbreak steffen.rebennack\}@kit.edu},\\
    \texttt{\{adrian.fessler, robert.lindermann\}@student.kit.edu},\\
  }
  \IEEEauthorblockA{
    \IEEEauthorrefmark{2}VINCI Energies, Basel, Switzerland\\
    \texttt{phil.arnold@vinci-energies.com},
    \texttt{andreas.hempel@actemium.de}
  }
}

\markboth{IEEE Transactions on Quantum Engineering}%
\IEEEpubid{}

\maketitle

\begin{abstract}
The Cable Routing Optimization Problem (CROP) is a Multi-Commodity Flow Problem (MCFP) central to industrial layouts and smart manufacturing. Historically, quantum optimization has modeled MCFPs as Quadratic Unconstrained Binary Optimization problems (QUBOs). Recent studies suggest that mapping routing problems to Polynomial Unconstrained Binary Optimization problems (PUBOs) can improve efficiency. However, solving full-scale MCFPs with quantum optimization remains computationally challenging. To bridge this gap, we introduce a Decomposed Variational Quantum Pipeline that exploits the block-diagonal structure of CROP, breaking the multi-cable routing task into modular, single-commodity subproblems. We explicitly derive both the QUBO and PUBO representations for CROP and demonstrate that our pipeline can evaluate both formulations within the same pipeline. Our empirical study highlights a trade-off: PUBO eliminates auxiliary qubits at the cost of circuit depth. In our experiments, the decomposed pipeline accelerates time-to-solution, reliably generating feasible cable layouts while trading strict optimality for computational scalability. PUBO formulations achieved full routing feasibility across all tested seeds, while global QUBO formulations showed substantially lower robustness.
\end{abstract}

\begin{IEEEkeywords}
Polynomial Unconstrained Binary Optimization, PUBO, Quadratic Unconstrained Binary Optimization, QUBO, Variational Quantum Algorithms, VQE, QAOA, Cable Routing Optimization

\end{IEEEkeywords}

\section{Introduction}

\subsection{Motivation for Cable Routing}
\IEEEPARstart{C}{able} routing is a critical task in modern engineering systems, from smart buildings to industrial plants. These systems require thousands of cables that must be routed through complex infrastructures of segments, conduits, and equipment. Designing efficient cable layouts is increasingly important as systems grow in scale and connectivity. Suboptimal cable routing in such contexts can drive up material usage, violate spacing standards, or complicate maintenance. Therefore, ensuring reliable electrical connectivity and cost-effective use of materials depends on cable layout decisions~\cite{Ittner2008}.

In practice, cable routing is still dominated by computer-aided design (CAD) tools and engineering heuristics. Experienced engineers manually plan routes, leveraging personal know-how to satisfy physical and regulatory constraints \cite{VanDerVelden2006, FrankRebennack2015}. The growing scale and complexity of cable systems have revealed the limitations of this traditional approach, motivating the search for optimization methods. 

\subsection{Practices and Challenges in Cable Routing Optimization}
The \textit{Cable Routing Optimization Problem (CROP)} can be formulated as a combinatorial optimization problem. Different types of cables (for instance medium- or low-voltage) must be physically routed through predefined pathways called \textit{segments}. The spatial arrangement and routing of cables can be treated as a graph of interconnected segments, where each segment represents a physical pathway between two connection points. The goal is to determine which segments should be used by each cable, such that all cables are routed from their \textit{source} to \textit{terminal} while minimizing total cable layout cost. \textit{Mixed-Integer Linear Programming (MILP)} formulations can then be utilized to capture the objective of minimizing total cable layout cost while enforcing connectivity and technical constraints \cite{Zhao2023}. However, the cable routing optimization problem becomes challenging to solve for industry-scale instances \cite{Ittner2008}. \\
Modeling approaches have been proposed for CROP. Cao et al.~\cite{cao2025mathematical} formulate a wind farm layout with integrated cable routing problem as a nonlinear problem, optimizing the cable routings between turbine positions. Results show that larger scale instances remain challenging to be solved with traditional optimization methods. In another line of work, Amin et al.~\cite{Amin2021SurveyMLRoutingSDN} survey ML approaches to cable routing problems and note a recurring limitation: complex and long-lasting training procedures. Routing optimization in various applications is challenging to solve, motivating the use of alternative computing paradigms.
\subsection{Quantum Computing and Routing Optimization}
Quantum computing (QC) is an emerging paradigm that harnesses the laws of quantum mechanics to solve combinatorial problems that challenge the limits of classical computing \cite{NielsenChuang2010, Bouchmal2024QAOA6G}. Building on this capability, Variational Quantum Algorithms (VQAs), such as the \textit{Variational Quantum Eigensolver (VQE)} \cite{Peruzzo2014} and \textit{Quantum Approximate Optimization Algorithm (QAOA)} \cite{Farhi2014}, execute short, parameterized quantum circuits. The parameters are tuned with classical optimization methods to bias the sampling distribution toward solutions with lower objective values, making QC a promising approach for combinatorial optimization \cite{Lin2025}. Historically, to apply these algorithms, an optimization problem had to be mapped to a \textit{Quadratic Unconstrained Binary Optimization problem (QUBO)}:
\begin{equation}
\min_{\mathbf{z}\in\{0,1\}^\ell} \ \mathbf{z}^\top Q\, \mathbf{z},
\label{eq:QUBO_general}
\end{equation}
where $\mathbf{z} = (z_i)_{i=1}^\ell$ is the $\ell$-dimensional binary decision vector and $Q\in\mathbb{R}^{\ell\times \ell}$ is a symmetric matrix. QUBOs were largely popularized by quantum algorithms, whose hardware only supports 2-local qubit interactions. However, a major drawback of QUBOs on gate-based devices is that enforcing complex constraints in the objective function often requires numerous auxiliary variables which increase the requirements for logical qubits \cite{stein2023evidencepubooutperformsqubo}. To avoid this qubit overhead, one can instead encode higher-order interactions directly at the Hamiltonian level. In gate-based devices, these interactions are typically compiled into sequences of one- and two-qubit gates, which reduces auxiliary-variable requirements but may increase circuit depth. Recent literature has shifted toward \textit{Polynomial Unconstrained Binary Optimization problems (PUBO)}:
\begin{equation}
\min_{\mathbf{z}\in\{0,1\}^\ell} \left( \sum_{i} c_i z_i + \sum_{i < j} c_{ij} z_i z_j + \sum_{i < j < k} c_{ijk} z_i z_j z_k + \dots \right),
\label{eq:PUBO_general}
\end{equation}
where $c \in \mathbb{R}$ represents the respective coefficients for linear, quadratic, and higher-order terms. By using polynomials of degree 3 or higher in \eqref{eq:PUBO_general}, those formulations can capture multi-variable couplings without relying on auxiliary variables. Recent studies suggest that, in some settings, PUBO formulations may yield more favorable optimization landscapes than QUBO formulations \cite{stein2023evidencepubooutperformsqubo}.

In the domain of routing, quantum algorithms have been investigated for problems such as vehicle routing \cite{Harwood2020RoutingQuantum} and 6G optical networks \cite{Bouchmal2024QAOA6G}. A recent study on the single-commodity Asset Retrieval Problem directly compared formulations of QUBO and PUBO, demonstrating that PUBO formulations reduced qubit requirements at the cost of deeper phase gadgets \cite{bell2025comparison}. Despite these advances, current literature is restricted to single-flow routing. Although Multi-Commodity Flow Problems (MCFP) have been explored via quantum algorithms, these efforts remain constrained to QUBO formulations \cite{Niu2025ApplyingQC}. 

\subsection{Contributions}
This paper investigates how quadratic and higher-order quantum optimization formulations can be used for MCFP in the context of cable routing. Through a decomposed pipeline, we explore the trade-offs of applying VQAs to CROP instances. Beyond modeling comparisons, we use the resulting QUBO and PUBO formulations to examine how auxiliary variables, higher-order interactions, decomposition, and feasibility filtering influence qubit requirements, runtime, and routing feasibility. This perspective is motivated by practical cable-layout workflows, where feasible layouts and scalable solution procedures are often more important than solving a fully coupled global formulation exactly~\cite{Ittner2008,VanDerVelden2006}. The main contributions are summarized as follows:
\begin{enumerate}
    \item \textbf{Comparative QUBO and PUBO Formulations for MCFP CROP.} We formalize CROP as a MCFP. For the undirected flow requirements, we explicitly derive both a QUBO, which relies on auxiliary variables, and a PUBO that eliminates auxiliary variable requirements. This comparison makes explicit how the same routing logic can lead to different quantum resource requirements depending on the chosen polynomial representation.
    \item \textbf{Decomposed Variational Pipeline with Empirical Analysis of Energy Landscapes.} To bypass intractable qubit scaling bottlenecks of global multi-flow routing, we introduce a hybrid pipeline that decomposes the MCFP into sequential single-commodity subproblems. We optimize these subproblems using different VQAs. This allows us to study hardware-relevant trade-offs: QUBO requires a larger number of qubits, whereas PUBO requires fewer qubits at the cost of deeper, more complex quantum circuits. The pipeline further demonstrates how inter-cable constraints can be handled through classical graph updates while keeping the quantum subproblems small.
    \item \textbf{Benchmark and Reproducible Evaluation Suite.} We will release a dataset derived from manufacturing layouts, including network topologies, heterogeneous multi-cable demands, and routing capacities. Alongside the dataset, we provide a software suite to build both QUBO and PUBO formulations for the CROP. This enables reproducible comparison of formulation choices, decomposition strategies, and VQA configurations for application-oriented cable routing experiments.
\end{enumerate}

\section{CROP}

\subsection{Mathematical model}
We model the CROP as a graph-based combinatorial optimization problem. Let $G = (K,S)$ be a finite, connected graph, where $K$ denotes the set of nodes and $S$ the set of segments. Each cable $c \in C$ is represented as a subgraph of $G$ connecting a source node $\hat{k}_c \in K$ and a terminal node $\check{k}_c \in K$. The set of all cables is denoted by $C$, which can be partitioned into disjoint subsets $C_{\mathrm{HV}}$ and $C_{\mathrm{LV}}$, i.e., $C = C_{\mathrm{HV}} \cup C_{\mathrm{LV}}$, where $C_{\mathrm{HV}}$ and $C_{\mathrm{LV}}$ represent the sets of high-voltage and low-voltage cables, respectively.

To represent the cables we use the binary variable $x_{cs}$ which is $1$ if cable $c$ lies in segment $s$ and $0$ otherwise. Let $\textbf{x}$ denote the binary vector with dimension $m = d\cdot n$ with $d= |S| $ and $n=|C|$. Additionally we have the cable segment costs $\alpha_{cs} >0$. Furthermore $\sigma(k)$ is the set of all segments incident to the node $k$. With that we can define $S_{c,c'}$ as the set of segments without segments incident to source and terminal nodes of the high- and low-voltage cables $c\in C_\text{HV}$ and $c'\in C_\text{LV}$, i.e.,
\begin{align*}
    S_{c,c'}= S \setminus (\sigma(\hat{k}_c) \cup \sigma(\check{k}_c) \cup \sigma(\hat{k}_{c'}) \cup \sigma(\check{k}_{c'})).
\end{align*}
Finally we define $\tilde{K}_c$ as the set of nodes without source and terminal node of a cable $c\in C$, i.e., $\tilde{K}_c = K\setminus \{ \hat{k}_c,\check{k}_c \} $.

\begin{align}
\min_{\textbf{x}\in\{0,1\}^m}
& \sum_{c \in C} \sum_{s \in S} \alpha_{cs} x_{cs} \label{obj:cable}\tag{i} \\
\text{s.t.} \quad 
& \sum_{s \in \sigma( \hat{k}_c) } x_{cs} = 1, \; \forall c\in C \label{con:cable_1} \tag{ii}\\
& \sum_{s \in \sigma( \check{k}_c) } x_{cs} = 1, \; \forall c\in C \label{con:cable_2} \tag{iii}\\
&\sum_{s\in \sigma(k)} x_{cs} \in \{ 0,2\} , \; \forall c\in C,\ \forall k \in \tilde{K}_c \label{con:cable_3} \tag{iv}\\
& \sum_{c \in C_{\text{HV}}} \sum_{c' \in C_{\text{LV}}} \sum_{s \in S_{c,c'}} x_{cs} x_{c's} = 0 \label{con:cable_4} \tag{v}
\end{align}
The objective \eqref{obj:cable} minimizes the weighted sum of a cost factor, i.e. cable length and cost.
Constraints \eqref{con:cable_1} and \eqref{con:cable_2} ensure the correct specification of the source and terminal nodes by enforcing that exactly one segment is incident to each of them for every cable. 
Constraint \eqref{con:cable_3} imposes flow conservation: for every node that is neither a source nor a terminal, an inflow exists if and only if there is a corresponding outflow. 
Thus, each such node either has no active segment or exactly two active segments for a given cable. 
Finally, constraint \eqref{con:cable_4} prevents high-voltage and low-voltage cables from occupying the same segment simultaneously.

\subsection{QUBO and PUBO formulations for CROP}
\subsubsection{Derivation of the QUBO Formulation}
We construct a QUBO formulation as in \eqref{eq:QUBO_general} by expressing the constraints as linear zero-equality conditions and incorporating them as quadratic penalty terms with weights $\lambda_1,\ldots,\lambda_4>0$, thereby penalizing infeasible solutions. As violations are integer-valued, the limitations of quadratic penalties are negligible in this setting.

Applying this approach to Constraint \eqref{con:cable_1} and \eqref{con:cable_2} yields the quadratic penalty terms
\begin{align*}
    P_{\text{src}} &= \sum_{c \in C} \Big( \sum_{s \in \sigma(\hat{k}_c)} x_{cs} - 1 \Big)^2,\\
    P_{\text{term}} &= \sum_{c \in C} \Big( \sum_{s \in \sigma(\check{k}_c)} x_{cs} - 1 \Big)^2.
\end{align*}


For the flow conservation constraint \eqref{con:cable_3} we introduce auxiliary binary variables $b_{ck} \in \{0,1\}$ to obtain a quadratic penalty after squaring. We define for each $c\in C$ and $k\in \tilde{K}_c$
\begin{align*}
    b_{ck} = \begin{cases}
        1, & \sum_{s\in\sigma(k)} x_{cs} \geq 1,\\
        0, & \text{otherwise}.
    \end{cases}
\end{align*}
So $b_{ck}=1$ indicates that node $k$ is active (i.e., incident to selected segments). The constraint can then be expressed via the quadratic penalty
\begin{equation*}
    P_{\text{flow}} = \sum_{c \in C} \sum_{k \in \tilde{K}_c} \Big( \sum_{s \in \sigma(k)} x_{cs} - 2\, b_{ck} \Big)^2.
    \label{eq:qubo_flow_penalty}
\end{equation*}


For the voltage class separation constraint \eqref{con:cable_4}, we use the fact that binary variables satisfy $x_{cs}^2 = x_{cs}$ for all $c\in C$ and $s\in S$, so that squaring the constraint yields the quadratic penalty
\begin{equation*}
    P_{\text{sep}} = \sum_{c \in C_{\text{HV}}} \sum_{c' \in C_{\text{LV}}} \sum_{s \in S_{c,c'}} x_{cs} x_{c's}.
    \label{eq:qubo_sep_penalty}
\end{equation*}


Then the QUBO objective is the sum of the objective \eqref{obj:cable}, denoted as $P_\text{obj}$, and the penalties given as
\begin{equation}
    H_{\text{QUBO}} = P_\text{obj} + \lambda_1 P_{\text{src}} + \lambda_2 P_{\text{term}} + \lambda_3 P_{\text{flow}} + \lambda_4 P_{\text{sep}}.
    \label{eq:qubo}
\end{equation}
In this QUBO formulation, the total number of binary variables is $m_q \coloneqq n(d + p)$. The variables are split between the physical routing segments ($nd$) and the node states ($np$). The function $H_{\text{QUBO}}$ from \eqref{eq:qubo} is a quadratic function and can therefore be formulated as \eqref{eq:QUBO_general} for $\textbf{z} = (\textbf{x}, \textbf{b})^{\top}$ with $\textbf{b}$ as the binary vector of $b_{ck}$.

\subsubsection{Derivation of the PUBO Formulation}
Similarly, we construct the PUBO objective function as in \eqref{eq:PUBO_general} but we do not need the linearity of the zero-equality conditions.

The penalty terms for constraints \eqref{con:cable_1}, \eqref{con:cable_2}, and \eqref{con:cable_4} are adopted unchanged from the QUBO formulation, as their linear structure did not require the introduction of auxiliary binary variables.

The remaining component is the flow constraint \eqref{con:cable_3} whereby we use that $a \in \{0,2 \} $ is equivalent to $a(a-2)=0$. Therefore we get
\begin{equation*}
    \tilde{P}_{\text{flow}} = \sum_{c \in C} \sum_{k \in \tilde{K}_c} \Big( \sum_{s \in \sigma(k)} x_{cs} \Big)^2 \Big( \sum_{s \in \sigma(k)} x_{cs} - 2 \Big)^2.
\end{equation*}

This yields the PUBO objective as
\begin{equation}
    H_{\text{PUBO}} = P_\text{obj} + \lambda_1 P_{\text{src}} + \lambda_2 P_{\text{term}} + \lambda_3 \tilde{P}_{\text{flow}} + \lambda_4 P_{\text{sep}}.
    \label{eq:pubo}
\end{equation}

By leveraging the inherent structure of the problem the PUBO formulation has the same dimension as before, $m_p \coloneqq m = d\cdot n$. The function $H_\text{PUBO}$ is a 4th-order polynomial and can therefore be formulated as \eqref{eq:PUBO_general} for $\textbf{z}=\textbf{x}$.

\section{Solving CROP with VQAs}

\subsection{Variational Quantum Algorithms}\label{sec:VQA}
VQAs, such as VQE and QAOA, are hybrid quantum-classical heuristics for approximately minimizing the energy of a problem Hamiltonian. The objective function is encoded in a Hamiltonian \(\hat H\), whose low-energy states may correspond to feasible or near-feasible low-cost solutions of the optimization problem. A parameterized quantum circuit prepares a trial state, its energy is estimated on a quantum processor, and a classical optimizer updates the circuit parameters $\boldsymbol{\theta}$, as illustrated in Fig.~\ref{fig:vqe-minimal-tight}.

\begin{figure}[ht]
\centering
\begin{tikzpicture}[
  x=1cm, y=0.8cm, >=Latex, line width=0.5pt, font=\small,
  every node/.style={inner sep=1.2pt, outer sep=0pt}
]

\def\Top{0.30}
\def\Bot{-2.70}

\foreach \y/\lbl in {0/{$|q_1\rangle$}, -0.8/{$|q_2\rangle$}, -1.6/{$\vdots$}, -2.4/{$|q_m\rangle$}}{
  \draw (0,\y) -- (4.6,\y);
  \node[left=2pt] at (0,\y) {\lbl};
}

\def\Uleft{1.0}
\def\Uright{2.2}
\pgfmathsetmacro{\Ucx}{(\Uleft+\Uright)/2}

\draw[fill=\colormapFiveFill, rounded corners=2pt] (\Uleft,\Top) rectangle (\Uright,-2.4);
\node at (\Ucx,-1.1) {$U(\boldsymbol{\theta})$};

\draw[densely dotted] (2.35,\Top) -- (2.35,-2.4);
\node[above] at (2.35,0.42) {$|\psi(\boldsymbol{\theta})\rangle$};

\draw[fill=\colormapFiveFill, rounded corners=2pt] (2.6,\Top) rectangle (4.6,-2.4);
\node at (3.6,-0.95) {Measure};
\node at (3.6,-1.55) {$E(\boldsymbol{\theta})$};

\coordinate (MeasRight) at (4.6,-1.1);
\node[draw, dashed, rounded corners=2pt, align=center,
      minimum width=2.0cm, minimum height=0.9cm,
      right=9mm of MeasRight] (OPT) {Classical\\ optimizer};

\draw[->] (MeasRight) -- (OPT.west);
\draw[->] (OPT.north) -- ($(OPT.north) + (0, 1)$);
\node[below=0pt, anchor=south, xshift=0pt] at ($(OPT.north) + (0, 1)$) {$E_{\exp}(\theta^{\star})$};
\node[above=2.2pt, yshift=-14pt, anchor=east, xshift=-3pt]
     at ($(OPT.north) + (0, 1)$) {\footnotesize Termination};

\coordinate (UbottomMid) at (\Ucx,-2.4);
\draw[->] (OPT.south) |- ($(UbottomMid)+(0,-0.6)$) -- (UbottomMid)
  node[pos=0.2, below=4pt] {update $\boldsymbol{\theta}$};

\end{tikzpicture}
\caption{VQA optimization loop. The parameterized quantum circuit \(U(\boldsymbol{\theta})\) prepares a state from which samples are measured. The classical optimizer evaluates the objective and updates \(\boldsymbol{\theta}\) until convergence.}
\label{fig:vqe-minimal-tight}
\end{figure}

The VQA objective can be expressed as \cite{Cerezo2021VQA}:
\begin{align}
    \boldsymbol{\theta}^{\star}
    =
    \underset{\boldsymbol{\theta}}{\operatorname{arg\,min}}\,
    C(\boldsymbol{\theta}),
    \qquad
    C(\boldsymbol{\theta})
    =
    \langle \psi(\boldsymbol{\theta})|
    \hat H
    |\psi(\boldsymbol{\theta})\rangle .
\end{align}
The specific form of \(|\psi(\boldsymbol{\theta})\rangle\) depends on the chosen ansatz. In the VQE setting, we use a hardware-efficient EfficientSU2 ansatz~\cite{qiskit_efficientsu2}. In contrast, QAOA uses a problem-specific ansatz that alternates between the problem Hamiltonian and a mixer Hamiltonian~\cite{Farhi2014}.

Repeated measurements of \(|\psi(\boldsymbol{\theta})\rangle\) induce a distribution \(p_{\boldsymbol{\theta}}(\mathbf{z})\) over bitstrings \(\mathbf{z}\in\{0,1\}^{m}\). The expected energy is estimated for \(H\in\{H_{\mathrm{QUBO}},H_{\mathrm{PUBO}}\}\) as
\begin{equation}
    E_{\mathrm{exp}}(\boldsymbol{\theta})
    =
    \sum_{\mathbf{z}}
    p_{\boldsymbol{\theta}}(\mathbf{z})\,H(\mathbf{z}),
    \label{eq:energy}
\end{equation}
and passed to a classical optimizer, such as COBYLA~\cite{powell1994direct}. Candidate routing solutions are then obtained from the measured bitstrings generated by the optimized circuit.

This distinction between VQE and QAOA is important for interpreting the QUBO--PUBO trade-off. QUBO formulations are standard in quantum optimization because they contain only quadratic, i.e., 2-local, interactions. However, in \eqref{eq:qubo} this comes at the cost of introducing auxiliary variables \(b_{ck}\), increasing the logical qubit count from \(m=nd\) in the PUBO case to \(m_q=n(d+p)\) in the QUBO case. The PUBO formulation in~\eqref{eq:pubo} avoids these \(np\) auxiliary variables and therefore reduces the number of logical variables, but the flow-conservation penalty \(\tilde{P}_{\mathrm{flow}}\) introduces fourth-order polynomial terms. For QAOA, these higher-order interactions must be compiled into multi-qubit phase gadgets, increasing circuit depth and sensitivity to noise~\cite{bell2025comparison}. Unlike QAOA, VQE with EfficientSU2 does not directly encode the problem Hamiltonian into circuit layers. It uses a generic parameterized circuit (no problem-specific circuit adjustment), while the Hamiltonian is only used to evaluate the energy of the measured bitstrings.Thus, PUBO reduces qubit requirements but may increase circuit complexity, particularly for QAOA, where higher-order terms must be compiled into multi-qubit operations. In contrast, VQE is less directly affected in circuit depth, as it relies on a hardware-efficient ansatz; however, higher-order Hamiltonians can still increase measurement overhead and variance in the energy estimation.

\subsection{The Decomposed Variational Pipeline}\label{sec:decomp_pipeline}
Instead of optimizing the fully-coupled multi-flow CROP in one global quantum problem, our pipeline routes cables sequentially via single-cable subproblems. The motivation is that \eqref{con:cable_1}--\eqref{con:cable_3} act locally for each cable, while the HV/LV separation constraint \eqref{con:cable_4} enforces inter-cable constraints. If \eqref{con:cable_4} is included in one global Hamiltonian, the problem has to represent all cables and their interactions at once, increasing both the required number of qubits and the circuit complexity.

For a given cable \(c\), the classical outer loop generates the corresponding single-cable Hamiltonian, either \(H_{\mathrm{QUBO},c}\) or \(H_{\mathrm{PUBO},c}\), and executes a VQA to obtain a candidate routing bitstring \(\hat{\mathbf{z}}_c\). After a feasible route is selected, the graph costs are updated before the next cable is routed. To enforce \eqref{con:cable_4}, segments already occupied by an LV cable are made unavailable for subsequent HV cables by assigning them a prohibitive routing cost, i.e., \(\alpha_{cs}\to\infty\), before the next Hamiltonian is generated. Thus, inter-cable constraints are enforced classically, while the quantum processor only solves single-commodity subproblems.

The fully-coupled CROP objective can be written schematically as
\[
\min_{\{\mathbf{z}_c\}_{c\in C}}
E(\{\mathbf{z}_c\}_{c\in C})
=
\sum_{c\in C} H_c(\mathbf{z}_c)
+
\sum_{\substack{c,c'\in C\\ c<c'}}
P_{cc'}(\mathbf{z}_c,\mathbf{z}_{c'}),
\]
where \(H_c(\mathbf{z}_c)\) contains the cable-local routing objective and constraints, while \(P_{cc'}(\mathbf{z}_c,\mathbf{z}_{c'})\) represents inter-cable constraints in \eqref{con:cable_4}. In the decomposed pipeline, the terms \(P_{cc'}\) are not loaded into one global quantum Hamiltonian. Instead, they are handled through sequential graph-cost updates.

Since routes are fixed one after another, early routing decisions can restrict the possible routes of cables routed afterwards. This corresponds to a greedy sequential approximation that may exclude globally optimal joint routing configurations. Thus, the method does not preserve the global optimality guarantee of the full CROP formulation. Its purpose is instead to reduce the size of the quantum problem and make VQA execution tractable for near-term problem instances.

\subsection{Algorithm Framework}
Algorithm~\ref{algo:decomp-vqe-e2e} summarizes the decomposed pipeline introduced in Section~\ref{sec:decomp_pipeline}. For each cable, a QUBO or PUBO Hamiltonian is generated using the current costs and penalties. After VQA convergence, we apply feasibility filtering~\cite{amaro2022filtering}: all unique sampled bitstrings are sorted by their evaluated classical energy, and the first candidate satisfying \eqref{con:cable_1}--\eqref{con:cable_4} is accepted as $\hat{\mathbf{z}}_c$. If no feasible bitstring is found, the run is classified as unsuccessful for that cable. The committed route is then used to update the graph costs before routing the next cable through $\texttt{Update\_Graph\_Costs}(\cdot,\cdot)$.

\begin{algorithm}[ht]
\begin{algorithmic}[1]
\Require Sequenced cables \(C\), segments \(S\), nodes \(K\), costs \(\boldsymbol{\alpha}\), penalties \(\boldsymbol{\lambda}\), generator \(F\in\{\texttt{Generate\_QUBO},\texttt{Generate\_PUBO}\}\), VQA parameters.
\Ensure Sequential routing assignment \(\mathbf{z}^{\star}\), total evaluated energy \(E_{\mathrm{total}}\)
\State Initialize \(\mathbf{z}^{\star}\leftarrow [\,]\), \(E_{\mathrm{total}}\leftarrow 0\)
\For{each cable \(c\in C\)}
    \State \(H_c \leftarrow F(c,S,K,\boldsymbol{\alpha},\boldsymbol{\lambda})\)
    \State \(\boldsymbol{\theta}\leftarrow \boldsymbol{\theta}_0\)
    \While{optimizer not converged}
        \State Estimate \(E_{\mathrm{exp}}(\boldsymbol{\theta})\) from \(U(\boldsymbol{\theta})\) and \(H_c\)
        \State \(\boldsymbol{\theta}\leftarrow \texttt{Classical\_Update}(\boldsymbol{\theta},E_{\mathrm{exp}})\)
    \EndWhile
    \State \(\hat{\mathbf{z}}_c \leftarrow\) lowest-energy feasible bitstring from \(p_{\boldsymbol{\theta}}(\mathbf{z}_c)\)
    \State \(E_c \leftarrow H_c(\hat{\mathbf{z}}_c)\)
    \State Append \(\hat{\mathbf{z}}_c\) to \(\mathbf{z}^{\star}\)
    \State \(E_{\mathrm{total}}\leftarrow E_{\mathrm{total}}+E_c\)
    \State \(\boldsymbol{\alpha}\leftarrow \texttt{Update\_Graph\_Costs}(\boldsymbol{\alpha},\hat{\mathbf{z}}_c)\)
\EndFor
\State \Return \(\mathbf{z}^{\star},E_{\mathrm{total}}\)
\end{algorithmic}
\caption{Decomposed Variational Routing Pipeline}
\label{algo:decomp-vqe-e2e}
\end{algorithm}

\section{Numerical Experiments}

\subsection{Experimental Setup}
The CROP Hamiltonians, the penalty weights $\lambda_1,\lambda_2,\lambda_3,\lambda_4$ for the source, terminal, and flow constraints are set to instance-dependent lower bounds where, 
\begin{align*}
    \lambda_1 &= 1 + \max_{c\in C}\;\sum_{s\in\sigma(\hat{k}_c)} \alpha_{cs},\; \lambda_2 = 1 + \max_{c\in C}\;\sum_{s\in\sigma(\check{k}_c)} \alpha_{cs},\\
    \lambda_3 &= 1 + \max_{c\in C}\;\max_{k\in\tilde{K}_c}\;\sum_{s\in\sigma(k)} \alpha_{cs},\; \lambda_4 = \bigl(1 + \max_{c \in C, s\in S}\;\alpha_{cs}\bigr). \label{eq:eta3}
\end{align*}
Each $\lambda_i$ is chosen to heuristically dominate the maximum local objective gain from a single constraint violation. The separation penalty $\lambda_4$ is chosen so that co-locating an HV and LV cable on any segment is penalized more strongly than the corresponding local routing benefit. In our experimental setup, we scale penalties with $2 \lambda_i$, for $i = 1,\dots,4$.

We evaluate the full and decomposed CROP formulations on a room-scale multi-commodity cable routing instance using these penalties. We consider a layout with $p=6$ nodes and $d=7$ segments on which $n=2$ cables must be routed simultaneously: one LV cable ($\vert C_{\text{LV}} \vert = 1$) and one HV cable ($\vert C_{\text{HV}} \vert = 1$), sharing a common source node. This yields four formulation approaches: Global QUBO ($22$ logical qubits), Global PUBO ($14$ logical qubits), Decomposed QUBO ($11$ qubits per sub-problem), and Decomposed PUBO ($7$ qubits per sub-problem).

Each problem is executed with both a QAOA and a VQE ansatz at depth $5$. For Algorithm~\ref{algo:decomp-vqe-e2e}, cables are routed sequentially: the LV cable is solved first, after which occupied segment costs are inflated by $\sum_{s \in S} \beta_s$ before solving the HV cable. Common solver settings unless stated otherwise are: \texttt{COBYLA} optimizer ($\rho_{\mathrm{beg}}=2.0$, $\mathrm{tol}=10^{-8}$), $\texttt{shots}=4096$ and $\texttt{maxiter}=3000$ for all configurations. 

To prevent the classical optimizer from being skewed by invalid and highly penalized routes, we use CVaR aggregation \cite{Barkoutsos2020improving}. With a CVaR threshold of 0.2, the objective is estimated from the lowest-energy 20\% of sampled bitstrings rather than from the full empirical mean. This biases the classical optimizer toward improving the best observed candidate routes. Each formulation approach with QAOA and VQE is performed for $15$ seeds and run on an AMD Ryzen 9 7900X @ 4.7\,GHz CPU.

\paragraph{Empirical probability of joint feasibility}
We compute the \textit{empirical probability of joint feasibility for approach $a$ with ansatz $\mathcal{A}$}, denoted $\operatorname{EmpProb}_{a}^{\mathcal{A}}$, over a set of random seeds $\mathcal{R}$, namely
\begin{equation*}
  \operatorname{EmpProb}_{a}^{\mathcal{A}}
    \coloneqq \frac{1}{|\mathcal{R}|}\sum_{r\in\mathcal{R}} f_{a,r}^{\mathcal{A}},
  \label{eq:empirical_prob}
\end{equation*}
where $f_{a,r}^{\mathcal{A}}=1$ if approach $a$ under ansatz $\mathcal{A}$ and seed $r$ returns a solution satisfying all constraints~\eqref{con:cable_1}--\eqref{con:cable_4} and $f_{a,r}^{\mathcal{A}}=0$ otherwise.

\paragraph{Relative Optimality Gap against Mean Wall Clock Time}
We evaluate the CROP objective~\eqref{obj:cable} and compare it to the optimum $\mathbf{x}^*=(\mathbf{x}_c^*)_{c\in C}$. We define the Relative Optimality Gap of approach $a$ with ansatz $\mathcal{A}$ per seed $r$ as
\begin{align*}
    \operatorname{OptGap}_{\text{rel}}^{a,\mathcal{A},r}
      \coloneqq \frac{\left|\,\displaystyle\sum_{c\in C}\alpha_c^\top \mathbf{z}_{c}^{a,\mathcal{A},r}
                       - \sum_{c\in C}\alpha_c^\top \mathbf{x}_{c}^*\right|}
                     {\left|\,\displaystyle\sum_{c\in C}\alpha_c^\top \mathbf{x}_{c}^*\right|},
\end{align*}
averaging over the feasible seeds $\mathcal{R}_{\text{feas}}^{a,\mathcal{A}}$ we obtain
\begin{align*}
    \operatorname{OptGap}_{\text{rel}}^{a,\mathcal{A}}
      \coloneqq \frac{1}{|\mathcal{R}_{\text{feas}}^{a,\mathcal{A}}|}
                \sum_{r\in\mathcal{R}_{\text{feas}}^{a,\mathcal{A}}}
                \operatorname{OptGap}_{\text{rel}}^{a,\mathcal{A},r},
\end{align*}
where $\mathcal{R}_{\text{feas}}^{a,\mathcal{A}}$ is the set of seeds that returned a solution satisfying all per-cable path constraints~\eqref{con:cable_1}--\eqref{con:cable_4}. We obtain the Mean Wall Clock Time by averaging across wall clock times across seeds. 

\paragraph{Discussion of Results}
As a baseline for generating optimal CROP solutions, we use Gurobi 12~\cite{gurobi}. Gurobi solves the CROP from \eqref{con:cable_1}--\eqref{con:cable_4}. On the considered instance, Gurobi requires approximately 0.005 seconds of wall-clock time. This runtime is not directly comparable to the reported quantum-optimization runtimes, since all VQA runs are simulated classically. Consequently, the reported VQA wall-clock times reflect classical simulation and optimizer overhead rather than physical quantum execution time, and are used to compare formulation and decomposition trade-offs rather than to claim computational advantage over classical solvers.

\begin{table}[h!]
    \caption{Empirical probability of joint feasibility across 15 seeds}
    \centering
    \begin{tabular}{|l|c|c|c|c|c|c|c|c|}
        \hline
        \raisebox{-0.10\totalheight}{Ansatz} & \textcolor{blue}{$\text{P}_\text{E}$} & \textcolor{blue}{$\text{Q}_\text{E}$} & \textcolor{blue}{$\text{P}_\text{A}$} & \textcolor{blue}{$\text{Q}_\text{A}$} & \textcolor{red}{$\text{P}_\text{E}$} & \textcolor{red}{$\text{Q}_\text{E}$} & \textcolor{red}{$\text{P}_\text{A}$} & \textcolor{red}{$\text{Q}_\text{A}$}\\
        \hline
        \raisebox{-0.05\totalheight}{EmpProb} & \raisebox{-0.10\totalheight}{$1.0$} & \raisebox{-0.10\totalheight}{$1.0$} & \raisebox{-0.10\totalheight}{$1.0$} & \raisebox{-0.10\totalheight}{$1.0$} & \raisebox{-0.10\totalheight}{$1.0$} & \raisebox{-0.10\totalheight}{$0.0$} & \raisebox{-0.10\totalheight}{$1.0$} & \raisebox{-0.10\totalheight}{$0.2667$}\\
        \hline
    \end{tabular}
    \label{tab:empirical_prob}
\end{table}
We denote by $\text{P}_\text{E}$ and $\text{P}_\text{A}$ the PUBO formulations solved with VQE and QAOA, respectively, and by $\text{Q}_\text{E}$ and $\text{Q}_\text{A}$ the corresponding QUBO formulations; decomposed variants are indicated in \textcolor{blue}{blue}, while global formulations are shown in \textcolor{red}{red}.

Table~\ref{tab:empirical_prob} reports the empirical feasibility probabilities across the considered ansätze. While all blue configurations achieve full feasibility, the red variants exhibit notable degradation, in particular $\textcolor{red}{\text{Q}_\text{E}}$ and $\textcolor{red}{\text{Q}_\text{A}}$, indicating increased sensitivity to the respective formulation. These findings already suggest structural differences in robustness, which are further reflected in the performance trade-offs discussed below.

\begin{figure}[h!]
\centering
\begin{tikzpicture}
\begin{groupplot}[
    group style={
        group size=1 by 2,   
        vertical sep=0.2cm
    },
    xmode=log,
    xmin=10,
    xmax=10000,
    width=8cm,
    height=5cm,
    legend style={font=\footnotesize,column sep=0.08cm, at={(0.0,1.05)},anchor=south west, nodes={scale=0.73, transform shape}},
    legend columns=2,
    legend image post style={scale=0.73},
    legend cell align=left,
    grid=both,
]

\nextgroupplot[ylabel={qubits},ytick={7, 11, 14, 22},xticklabel=\empty, ymin=4, ymax=25]
\addplot[only marks, mark=text, text mark={$\text{P}_\text{E}$}, color=blue]
coordinates {(5000000, 50)};
\node[circle, fill, blue, inner sep=1pt, label={[xshift=-3pt,yshift=-3pt, color=blue]above right:$\text{P}_\text{E}$}] at (265.2, 7) {};

\addplot[only marks, mark=text, text mark={$\text{P}_\text{E}$}, color=red]
coordinates {(5000000, 50)};
\node[circle, fill, red, inner sep=1pt, label={[xshift=-3pt,yshift=-3pt, color=red]above right:$\text{P}_\text{E}$}] at (1306.19, 14) {};

\addplot[only marks, mark=text, text mark={$\text{Q}_\text{E}$}, color=blue]
coordinates {(5000000, 50)};
\node[circle, fill, blue, inner sep=1pt, label={[xshift=-3pt,yshift=-3pt, color=blue]above right:$\text{Q}_\text{E}$}] at (970.51, 11) {};

\addplot[only marks, mark=text, text mark={$\text{Q}_\text{E}$}, color=red]
coordinates {(5000000, 50)};
\node[circle, fill, red, inner sep=1pt, label={[xshift=-3pt,yshift=-3pt, color=red]above right:$\text{Q}_\text{E}$}] at (5561.49, 22) {};

\addplot[only marks, mark=text, text mark={$\text{P}_\text{A}$}, color=blue]
coordinates {(5000000, 50)};
\node[circle, fill, blue, inner sep=1pt, label={[xshift=-3pt,yshift=-3pt, color=blue]above right:$\text{P}_\text{A}$}] at (46.23, 7) {};

\addplot[only marks, mark=text, text mark={$\text{P}_\text{A}$}, color=red]
coordinates {(5000000, 50)};
\node[circle, fill, red, inner sep=1pt, label={[xshift=-3pt,yshift=-3pt, color=red]above right:$\text{P}_\text{A}$}] at (149.87, 14) {};

\addplot[only marks, mark=text, text mark={$\text{Q}_\text{A}$}, color=blue]
coordinates {(5000000, 50)};
\node[circle, fill, blue, inner sep=1pt, label={[xshift=-3pt,yshift=-3pt, color=blue]above right:$\text{Q}_\text{A}$}] at (119.86, 11) {};


\addplot[only marks, mark=text, text mark={$\text{Q}_\text{A}$}, color=red]
coordinates {(5000000, 50)};
\node[circle, fill, red, inner sep=1pt, label={[xshift=-3pt,yshift=-3pt, color=red]above right:$\text{Q}_\text{A}$}] at (1549.07, 22) {};

\legend{Decomposed PUBO with VQE, Global PUBO with VQE, Decomposed QUBO with VQE, Global QUBO with VQE, Decomposed PUBO with QAOA, Global PUBO with QAOA, Decomposed QUBO with QAOA, Global QUBO with QAOA}

\nextgroupplot[ymode=log, ymin=0, ymax=2, xlabel style={font=\small}, xlabel={Mean Wall Clock Time in seconds}, ylabel={$\text{OptGap}_\text{rel}$}]
\addplot[opacity=0,only marks, mark=text, text mark={$\text{P}_\text{E}$}, color=blue]
coordinates {(265.2, 0.3646098431373483)};
\node[circle, fill, blue, inner sep=1pt, label={[xshift=-3pt,yshift=-3pt, color=blue]above right:$\text{P}_\text{E}$}] at (265.2, 0.3646098431373483) {};

\addplot[opacity=0,only marks, mark=text, text mark={$\text{P}_\text{A}$}, color=blue]
coordinates {(46.23, 0.3650124945740457)};
\node[circle, fill, blue, inner sep=1pt, label={[xshift=-3pt,yshift=-3pt, color=blue]above right:$\text{P}_\text{A}$}] at (46.23, 0.3650124945740457) {};

\addplot[opacity=0,only marks, mark=text, text mark={$\text{Q}_\text{E}$}, color=blue]
coordinates {(970.51, 0.3469773711585817)};
\node[circle, fill, blue, inner sep=1pt, label={[xshift=-3pt,yshift=-3pt, color=blue]above right:$\text{Q}_\text{E}$}] at (970.51, 0.3469773711585817) {};

\addplot[opacity=0,only marks, mark=text, text mark={$\text{Q}_\text{A}$}, color=blue]
coordinates {(119.86, 0.37329795121789217)};
\node[circle, fill, blue, inner sep=1pt, label={[xshift=-3pt,yshift=-3pt, color=blue]above right:$\text{Q}_\text{A}$}] at (119.86, 0.37329795121789217) {};

\addplot[opacity=0,only marks, mark=text, text mark={$\text{P}_\text{E}$}, color=red]
coordinates {(1306.19, 0.024307322875823222)};
\node[circle, fill, red, inner sep=1pt, label={[xshift=-3pt,yshift=-3pt, color=red]above right:$\text{P}_\text{E}$}] at (1306.19, 0.024307322875823222) {};

\addplot[opacity=0,only marks, mark=text, text mark={$\text{P}_\text{A}$}, color=red]
coordinates {(149.87, 0.09763194293999035)};
\node[circle, fill, red, inner sep=1pt, label={[xshift=-3pt,yshift=-3pt, color=red]above right:$\text{P}_\text{A}$}] at (149.87, 0.09763194293999035) {};

\addplot[opacity=0,only marks, mark=text, text mark={$\text{Q}_\text{A}$}, color=red]
coordinates {(1549.07, 0.14699448590480096)};
\node[circle, fill, red, inner sep=1pt, label={[xshift=-3pt,yshift=-3pt, color=red]above right:$\text{Q}_\text{A}$}] at (1549.07, 0.14699448590480096) {};

\end{groupplot}
\end{tikzpicture}
\caption{$\operatorname{OptGap}_{\text{rel}}^{a,\mathcal{A}}$ versus mean wall-clock time for global and decomposed QUBO and PUBO formulations combined with VQE and QAOA (bottom). Corresponding qubit counts for each configuration are shown in the top plot.}
\label{fig:result_plot}
\end{figure}

The results in Fig.~\ref{fig:result_plot} indicate a trade-off between solution quality, runtime, and qubit requirements. Decomposed approaches consistently reduce qubit counts and achieve lower wall-clock times compared to their global counterparts. In contrast, global formulations suffer from scalability limitations, particularly for the QUBO, where increased qubit requirements due to auxiliary variables $b_{ck}$ negatively impact both feasibility and stability.

While the decomposed PUBO approaches (\textcolor{blue}{$\text{P}_\text{A}$, $\text{P}_\text{E}$}) reduce wall-clock time in our experiments compared to their global counterparts, this improvement comes at the cost of slightly higher relative optimality gaps. In contrast, the global PUBO with VQE (\textcolor{red}{$\text{P}_\text{E}$}) achieves the smallest observed optimality gap, approximately 2\%, but requires higher runtime, highlighting a clear trade-off between solution quality and computational effort within PUBO formulations. This behavior reflects the underlying hardware trade-off introduced earlier: although PUBO formulations eliminate auxiliary variables and reduce qubit requirements, the resulting higher-order terms increase circuit depth and susceptibility to noise. 

The decomposed setting leads to faster runtimes but degradation in solution quality, whereas the global PUBO configuration better exploits the energy landscape at the cost of increased computational effort. However, the decomposed PUBO formulations still reliably produce feasible routing solutions across seeds, making them a practical alternative in settings where computational efficiency is critical.

Overall, on the considered room-scale instance, the results indicate that QUBO benefits from shallower circuit representations but suffers from increased qubit requirements due to auxiliary variables, while PUBO reduces qubit counts at the expense of deeper, more noise-sensitive circuits. For the tested instance, the benefit of qubit reduction in PUBO configurations appears to outweigh the cost of increased gate complexity, both in terms of runtime and solution optimality (contrary to the concerns raised in Section~\ref{sec:VQA}). Furthermore, decomposition is important to mitigate qubit limitations, enabling tractable optimization while exposing the fundamental balance between qubit efficiency, circuit complexity, and solution quality.

\section{Conclusions and Limitations}\label{sec-conclusions}
This paper presented a decomposed variational quantum optimization pipeline for CROP, motivated by an industrial cable-layout setting. Instead of encoding the full multi-cable problem as a single optimization problem, we exploit the partial cable-wise structure of CROP and solve sequential single-cable subproblems. Inter-cable conflicts, are handled through classical graph-cost updates. 

Our results show that formulation choice is a hardware-relevant engineering decision. Based on the benchmark data, QUBO keeps the Hamiltonian quadratic, but requires auxiliary variables for undirected flow constraints and therefore increases the logical qubit count. PUBO removes these variables and reduces qubit requirements, but introduces higher-order terms that may increase circuit depth, especially for QAOA. For the tested instance, PUBO-based formulations achieved full routing feasibility across seeds, while global QUBO was less robust. Global PUBO with VQE obtained the smallest observed optimality gaps, whereas decomposed PUBO offered faster execution with slightly larger gaps.

In our experimental setup, decomposition is important for obtaining feasible routes with reduced computational effort. Decomposed formulations consistently lowered qubit counts and wall-clock times compared to their global counterparts. This comes at the cost of losing global optimality guarantees, since routes are fixed sequentially. From an engineering perspective, this trade-off is shows that the pipeline reliably produced feasible cable layouts while reducing computational burden, making it a practical candidate for early-stage design exploration and quantum-assisted cable routing workflows.

Some limitations remain. The experiments use a room-scale instance and a classical simulator, so wall-clock times reflect simulation and optimization overhead rather than physical quantum execution time. The results also depend on penalty scaling and ansatz choice which may require rigorous tuning procedures. Finally, the current model captures a core subset of cable-routing constraints, while industrial layouts may require additional geometric, regulatory, and maintenance constraints. Future work will focus on larger industrial layouts, hardware deployment, and the integration of additional constraints relevant in large scale industrial settings.

\clearpage

\appendices

\bibliographystyle{ieeetr}
\bibliography{Manuscript_qce_26.bib}

\end{document}